\documentclass[prl,twocolumn,a4paper,amsmath,floatfix]{revtex4}
\usepackage[utf8]{inputenc}
\usepackage{graphicx}
\usepackage{amsmath}

\usepackage{color}
\newcommand{\ntet}{n_\mathrm{tet}}

\begin{document}

\title{Tetrahedrality dictates dynamics in hard spheres}

\author{Susana Marín-Aguilar}
\affiliation{Laboratoire de Physique des Solides, CNRS, Universit\'e Paris-Sud, Universit\'e Paris-Saclay, 91405 Orsay, France}
\author{Henricus H. Wensink}
\affiliation{Laboratoire de Physique des Solides, CNRS, Universit\'e Paris-Sud, Universit\'e Paris-Saclay, 91405 Orsay, France}
\author{Giuseppe Foffi }
\affiliation{Laboratoire de Physique des Solides, CNRS, Universit\'e Paris-Sud, Universit\'e Paris-Saclay, 91405 Orsay, France}
\author{Frank Smallenburg }
\affiliation{Laboratoire de Physique des Solides, CNRS, Universit\'e Paris-Sud, Universit\'e Paris-Saclay, 91405 Orsay, France}


\maketitle

{\bf
Glasses are ubiquitous amorphous solids that remain one of the big mysteries in condensed matter. Despite the vast body of literature on glasses, a unifying approach to link the structure and dynamics of glasses is still missing\cite{berthier2011theoretical,stillinger2013glass}. A growing set of evidence, however, indicates the microscopic local geometry as a key ingredient\cite{royall2015role,tanaka2019revealing}. This originated from the seminal work of Frank \cite{frank1952supercooling}, who conjectured that glasses may be the result of the local tendency of liquids to form icosahedral structures, which are not capable of globally filling space regularly. Here, we show that, for a fundamental glass model, dynamics can be fully understood by simply counting the number of tetrahedra. Both globally and locally, these local structures directly predict dynamical slowdown. After more than 60 years of Frank’s Conjecture, it might not be the icosahedra that matter for glasses, but rather the  tetrahedra inside them.}



\flushbottom




The elusiveness of the glass transition arises from the lack of an accepted theoretical framework. In general, two main approaches  have emerged \cite{tanaka2019revealing}. On a global level, different theoretical approaches have been unified under the random first order phase transition (RFOT) paradigm~\cite{kirkpatrick1989scaling,bouchaud2004adam}. This is essentially built on the original idea of Adam and Gibbs \cite{adam1965temperature} who hypothesized the existence of large cooperative rearranging regions that grow in size when the system gets deeper into the glassy regime. On a local geometrical level, Frank’s conjecture  \cite{frank1952supercooling} has been a great source of inspiration and different observables have been proposed to try to pin down the relation between local geometry and the extremely long timescales observed in glasses.  In particular, several authors have looked at the role of locally favored clusters with perfect or modified icosahedral structures as well as, more recently, at the local packing capability \cite{tanaka2019revealing}. For comprehensive comparisons of most of these methods, we suggest the recent reviews in Refs. \onlinecite{royall2015role} and Ref. \onlinecite{tanaka2019revealing}. In general, the emerging picture is that icosahedra and closely related clusters are long-lived structures that correlate strongly with the slow dynamics in glasses.

In this Letter, we examine the link between local structure and dynamics using simulations of hard-sphere mixtures. Hard spheres are a prototypical model glass former \cite{bengtzelius1984dynamics,barrat1989liquid, pusey2009hard, parisi2010mean, sanz2014avalanches}, and in mixtures their dynamics can be tuned not only via the packing fraction, but also by varying  the size ratio and composition \cite{gotze2003effect, foffi2003mixing}. Moreover, hard spheres can be perfectly reproduced in colloidal experiments, which have become a standard testing ground for glass theory \cite{pusey1987observation, van1993glass, weeks2000three, berthier2005direct, brambilla2009probing, kegel2000direct}. As such, hard-sphere mixtures are an ideal starting point for testing new ways to link structure and glassy dynamics. Here, we quantify local order via the tetrahedrality of the local structure (TLS), a simple idea which consists of counting the number of tetrahedral clusters around each particle. Various order parameters based on local (poly)tetrahedral order have previously been shown to correlate with glassy dynamics in soft spheres \cite{tong2018revealing}, granular systems \cite{xia2015structural}, and hard spheres \cite{anikeenko2007polytetrahedral, charbonneau2012geometrical}. Our results show that TLS not only has an impressively strong correlation with local dynamics, but is also able to quantitatively predict the diffusivity of a vast range of dense hard-sphere mixtures. As perfect icosahedra can be broken down into tetrahedra, TLS is a natural extension of Frank’s idea, but focuses on the most elementary building block of the three-dimensional fluid. Hence, we argue that tetrahedra are the slow structures that are ultimately responsible for dynamic arrest.

We begin our study by exploring the structure and dynamics of dense binary hard-sphere mixtures, varying the size ratio $q = \sigma_{S}/\sigma_{L}$ (with $\sigma_{S (L)}$ the diameter of the small (large) spheres), and the composition $x_L$, which denotes the fraction of large spheres. For each choice of $q$ and $x_L$, we prepare the system at a packing fraction of $\eta = 0.575$, allow it to equilibrate and measure the diffusion time $\tau_D$, defined as the time it takes a particle to diffuse over a distance $\sigma_L$. Note that we only consider systems which avoid crystallization; as a result, we cannot probe compositions close to 0 or 1, or size ratios close to $q=1$. 

\begin{figure}
\begin{center}
    \includegraphics[width=0.9\linewidth]{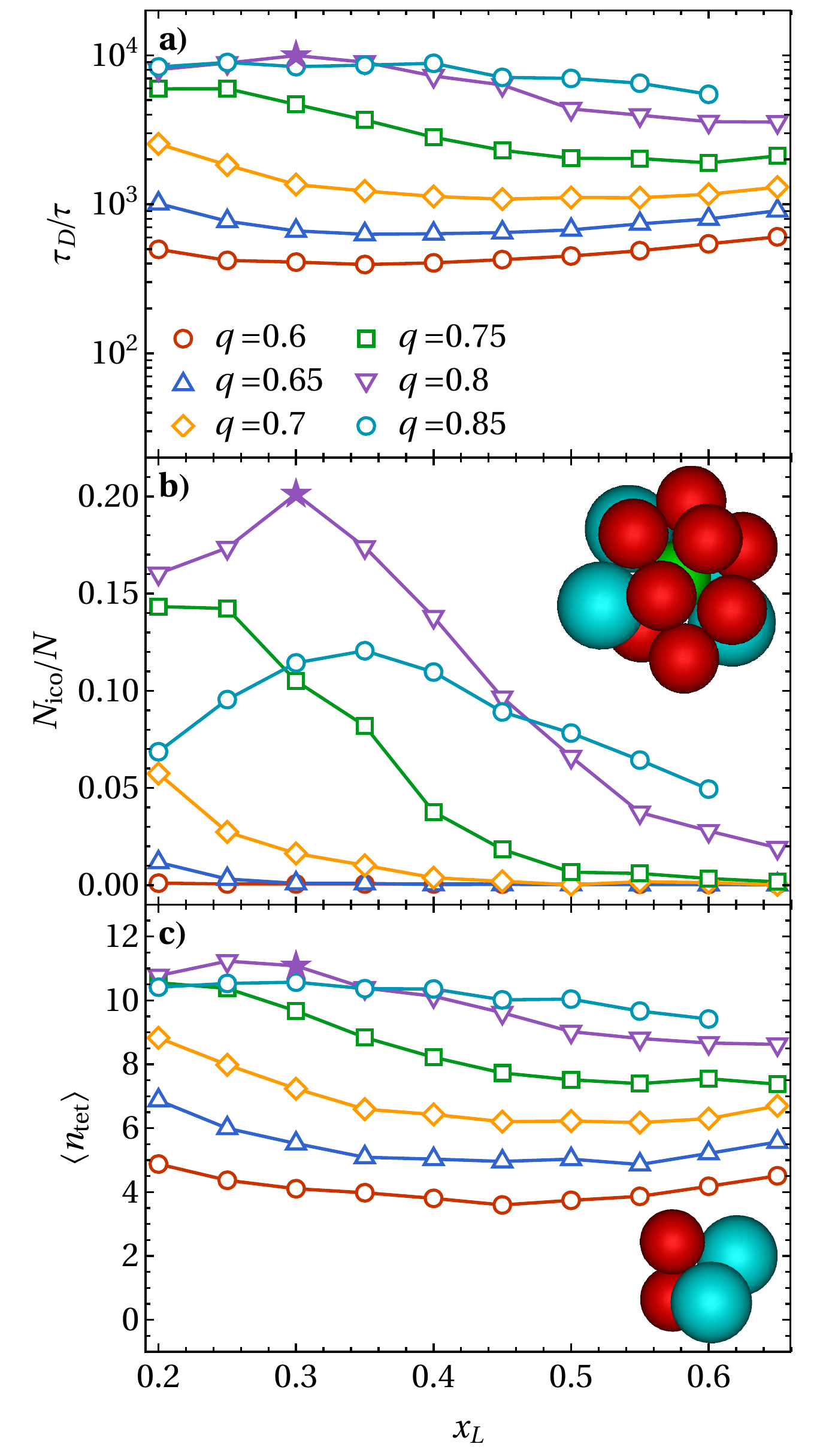}   
\end{center}
\caption{a) Diffusion time $\tau_D$ as a function of composition $x_L$ for binary hard-sphere mixtures with various size ratios $q$ as indicated. $\tau$ is the time unit of the simulation. The star indicates the point corresponding to the analysis in Fig. \ref{fig:localcorrelation} b) Fraction of particles inside an icosahedral cluster for the same systems. The inset shows a typical icosahedral cluster. c) Average number of tetrahedra per particle. The inset shows a typical tetrahedral cluster.}
\label{fig:binary_results}
\end{figure}

The dynamics of binary hard-sphere mixtures are known to vary significantly upon changing the composition $x_L$ and size ratio $q$ \cite{foffi2003mixing, gotze2003effect}. Our systematic investigation confirms this observation, as shown in Fig. \ref{fig:binary_results}a where we plot the diffusion time $\tau_D$ as a function of $x_L$ for different size ratios $q$. Intriguingly, even though the packing fractions of all systems are the same, $\tau_D$ varies by more than an order of magnitude. Moreover, the dependence of $\tau_D$ on $x_L$ is non-monotonous, and shows qualitatively different behavior for different size ratios. For mixtures with a small size ratio $q \lesssim 0.75$, the diffusion time is a convex function of $x_L$, showing a clear single minimum. In contrast, for higher size ratios, a maximum in the diffusion times appears, corresponding to some of the slowest systems we investigated.

The composition and size ratio of a hard-sphere mixture control the geometry of local packings in the fluid. Hence, the strong variation in diffusivity in Fig. \ref{fig:binary_results}a hints at strong variations in the local structure of these different systems as well. As local icosahedral order is known to slow down dynamics \cite{frank1952supercooling, royall2015role, shortpaper}, we measure the fraction of particles involved in at least one local icosahedral cluster (as determined via the Topological Cluster Classification algorithm \cite{malins2013tcc}). As shown in Fig. \ref{fig:binary_results}b, the fraction of particles in an icosahedral environment indeed shows behavior similar to that of the diffusion time. In particular, the slowest systems in our study (those with $q \simeq 0.8$ and $x_L \simeq 0.3$, marked by a star in Fig. \ref{fig:binary_results}a) have the highest degree of icosahedral order. However, Fig. \ref{fig:binary_results}b clearly does not reproduce all of the behavior of the diffusion time in Fig. \ref{fig:binary_results}a, and hence it is unlikely that icosahedra can be seen as the only slow structural motif in the system. Indeed, recent work has shown that a variety of complex clusters in hard-sphere mixtures can have long lifetimes \cite{malins2013identification}, and searching for these motifs in our systems reveals a whole family of clusters that commonly appear together with perfect icosahedral clusters (see SI). Intuitively, it is therefore likely that each of these contributes, to some degree, to the slowdown of the system. 

Interestingly, one feature shared by all of these local structural motifs is their incorporation of a large number of smaller tetrahedral clusters: groups of four particles in which each pair are considered nearest neighbors, based on a modified Voronoi construction (see Methods). Hence, a natural question to ask is whether the overall tetrahedrality of the local structure (TLS) can predict dynamical behavior. As essentially all particles are involved in multiple tetrahedral clusters, we quantify tetrahedrality by measuring the average number $\ntet$ of tetrahedra a particle is involved in (see Methods), and plot the results in Fig. \ref{fig:binary_results}c for each binary mixture. This simple structural order parameter captures the behavior of the diffusion time almost perfectly, reproducing both the convexity of $\tau_D$ for low $q$ and its maximum at high $q$.

 \begin{figure}
\begin{center}
    \includegraphics[width=0.98 \linewidth]{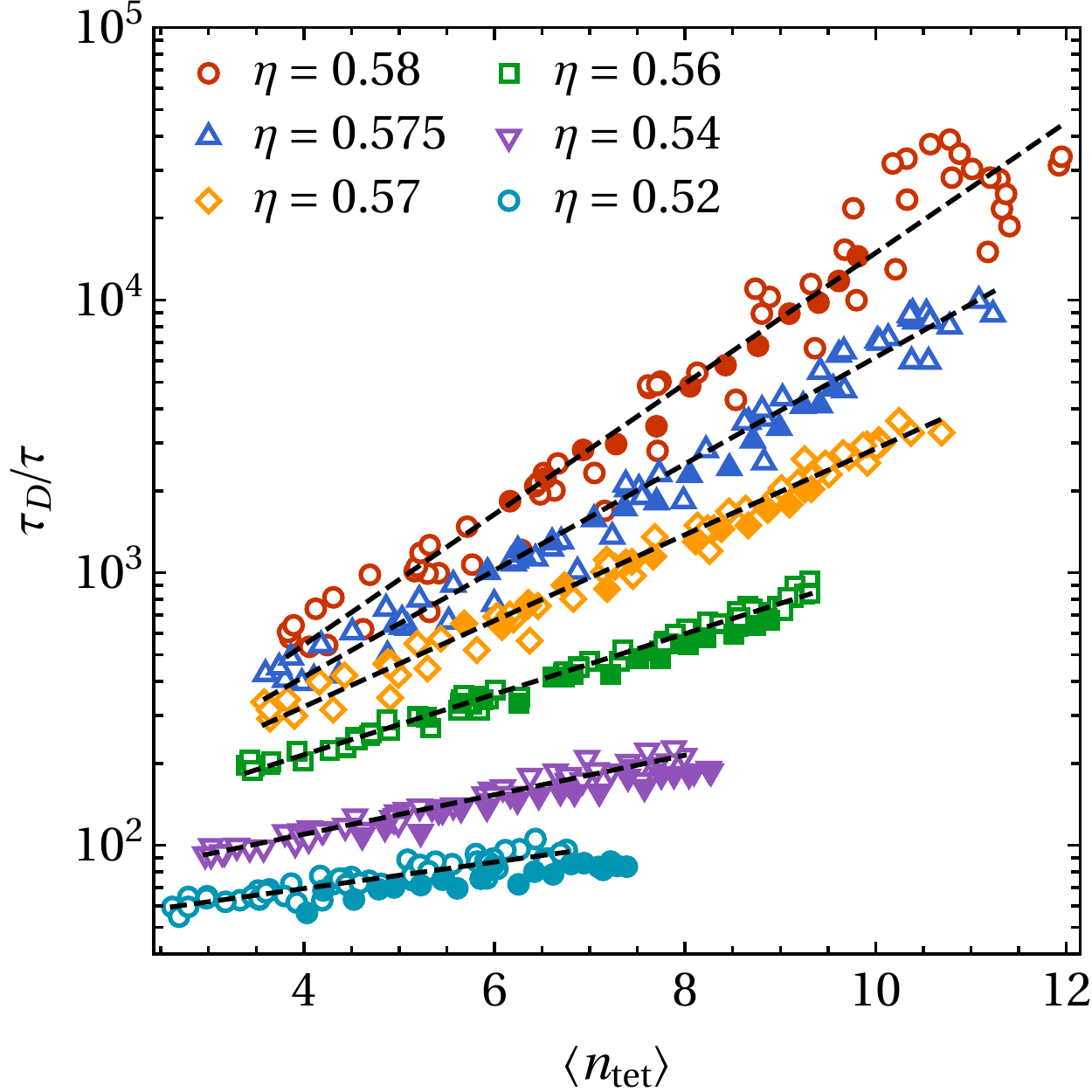}   \\
\end{center}
\caption{Diffusion time for all investigated hard-sphere mixtures as a function of tetrahedrality. Different colors indicate different packing fractions. Within each packing fraction, open symbols correspond to binary mixtures with different size ratios and compositions. Closed symbols are polydisperse systems with different packing fractions. The dashed lines are exponential fits to the binary data for each packing fraction. }
\label{fig:diffusiontime}
\end{figure}

The strong correlation between tetrahedrality and diffusivity becomes even more clear when we plot the relationship between these two quantities directly. In Fig. \ref{fig:diffusiontime}, we plot  $\tau_D$ vs $\langle \ntet \rangle$ for all investigated size ratios and compositions, and for a range of packing fractions. For each  packing fraction, $\ntet$ provides an excellent predictor of the diffusion time, revealing an approximately exponential relationship between $\tau_D$ and $\langle \ntet \rangle$ (dashed lines). Note that just the set of blue triangles in Fig. \ref{fig:diffusiontime} covers {\em all} 60 different systems in Fig. \ref{fig:binary_results}, with vastly different size ratios and compositions. Moreover, this data collapse is not restricted to binary systems: polydisperse systems fall on the same lines for all investigated polydispersities, ranging from 1\% to 20\% (closed symbols in Fig. \ref{fig:diffusiontime}).  This strongly suggests that this behavior is universal for all mixtures of hard spheres, with the possible exception of extreme size ratios where depletion effects could start to play a role \cite{anderson2002insights}.

\begin{figure}[t!]
\begin{center}
\includegraphics[width=0.85\linewidth]{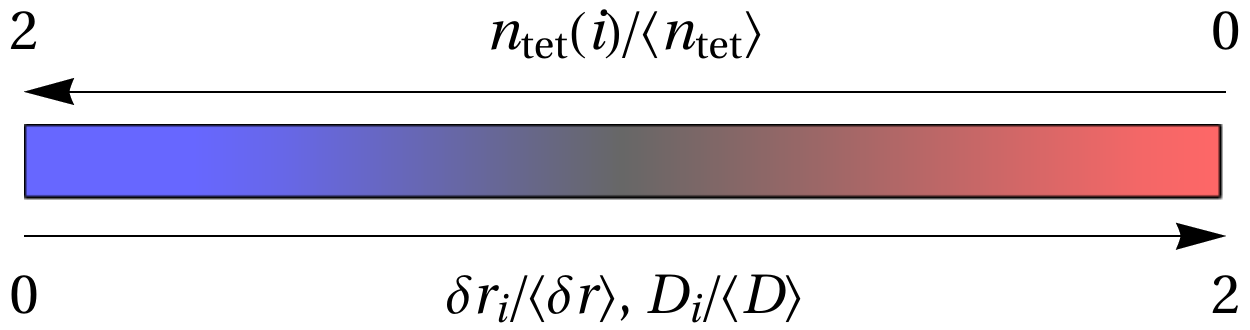}
    \begin{tabular}{cc}
        a)  $\ntet(i)$ & b)  $\delta r_i$\\
        \includegraphics[width=0.475\linewidth]{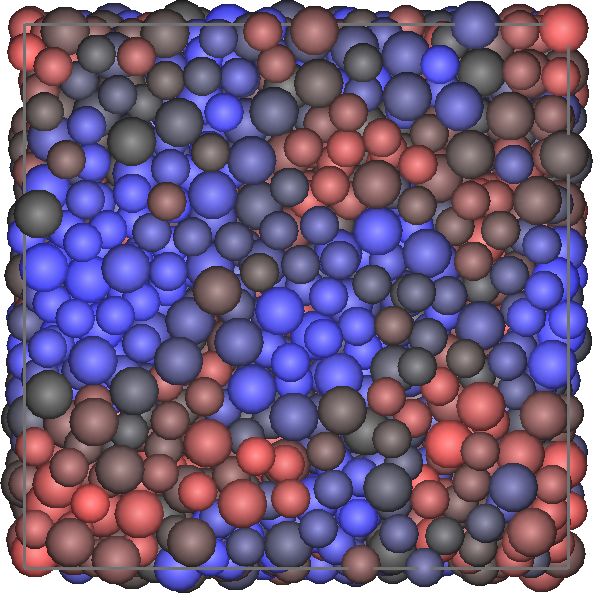}   &
        \includegraphics[width=0.475 \linewidth]{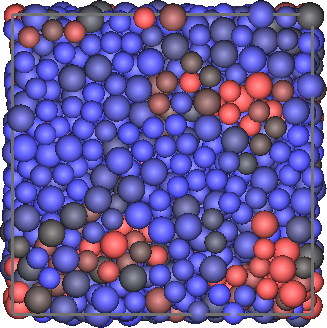}   \\
        c) $D_i$   &  d)\\
        \includegraphics[width=0.475\linewidth]{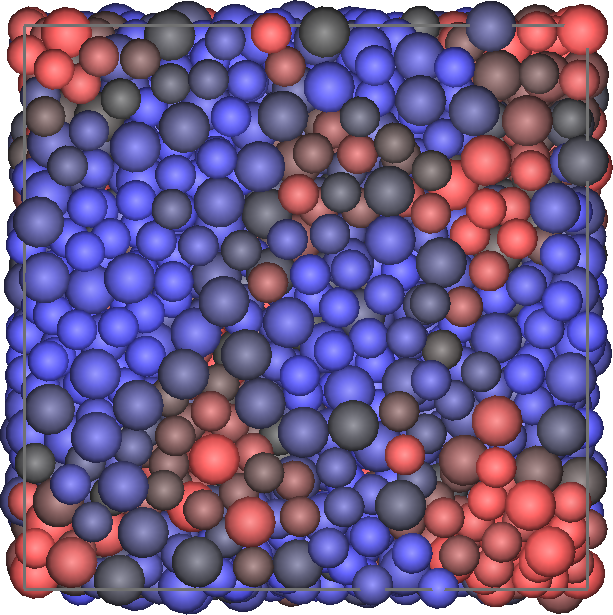}   &
        \includegraphics[width=0.475 \linewidth]{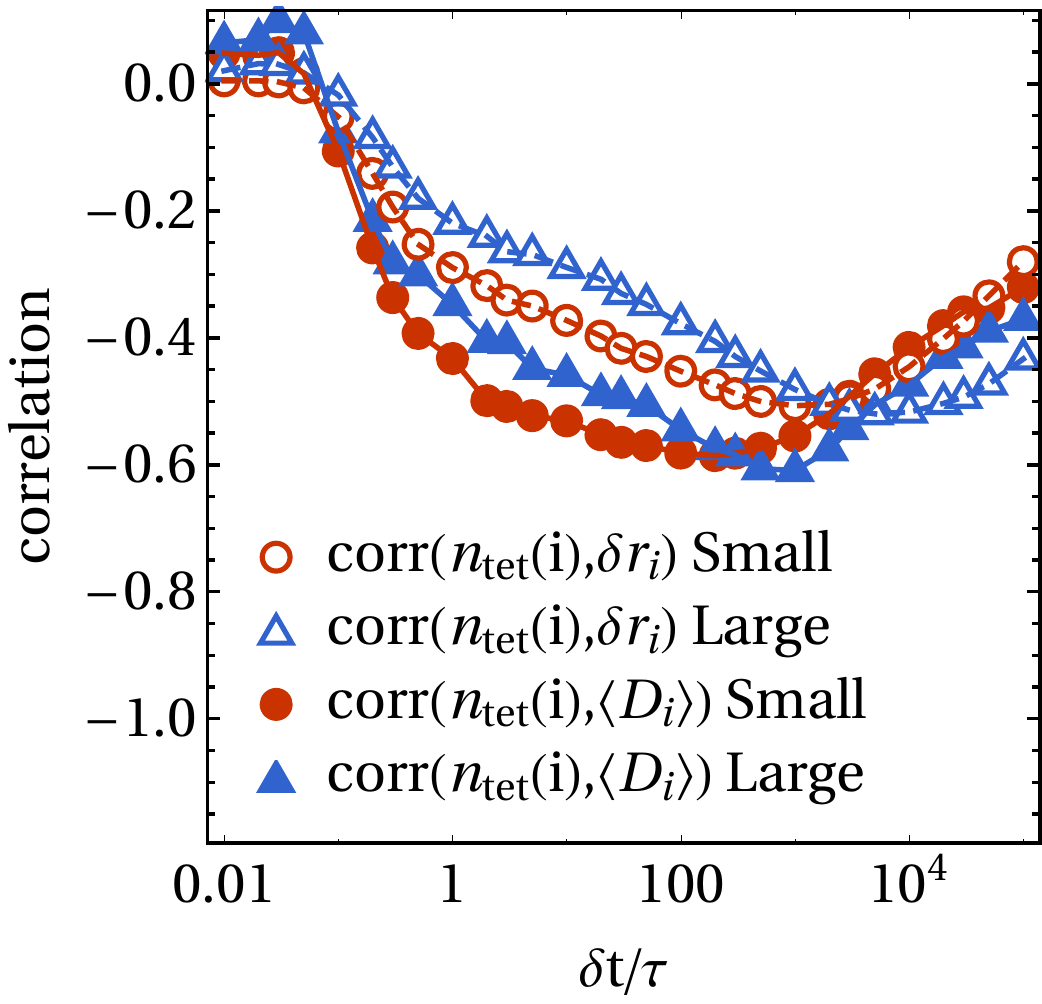}           
    \end{tabular}\\
\end{center}
\caption{a,b,c) Snapshot of a glassy system at packing fraction $\eta = 0.575$, size ratio $q = 0.8$ and composition $x_L = 0.3$ (denoted by a star in Fig. \ref{fig:binary_results}), with particles colored according to different criteria. a) According to the number of tetrahedra $\ntet(i)$ a particle is involved in, with red particles involved in fewer tetrahedra, and blue particles in more.  b) According to the absolute displacement $\delta r_i$ after a time interval $\delta t = 500 \tau$ in a random trajectory, with red indicating fast particles and blue indicating slow ones. c) According to the dynamic propensity $D_i$ over the same time interval.  In all snapshots, the color gray indicates the average for both large and small particles. c) Spearman's rank correlation between the number of tetrahedral clusters a particle is involved in $\ntet$ and either its displacement $\delta r_i$ (open symbols) or its dynamic propensity $D_i$ (closed symbols), measured over a time interval $\delta t$.}
\label{fig:localcorrelation}
\end{figure}

Thus far we have examined the relationship between globally averaged TLS and diffusivity. We now turn our attention to the impact of tetrahedral clusters on dynamics at a local level. 
To this end, we explore the relationship between the number of tetrahedra $\ntet(i)$ a given particle $i$ is involved in, and the absolute distance $\delta r_i$ over which it moves in a given time interval $\delta t$. In Fig. \ref{fig:localcorrelation}a and b, we show a typical snapshot of one of our slowest systems, with particles colored by either their tetrahedrality (a) or their displacement after a time interval $\delta t = 500 \tau$ in a typical trajectory (b). There is clear evidence of correlation between the two quantities, with regions of low tetrahedrality matching regions of high mobility (red), and vice versa (blue). However, examining the displacement in a specific trajectory provides only a limited view of particle mobility. After all, in a given trajectory, the ability of a particle to move depends not only on its environment, but also on the initial velocities of all particles. To average out this thermal noise, we measure the dynamic propensity $D_i$ of a particle: its average absolute displacement, taken over an ensemble of simulations starting from the same initial configuration \cite{widmer2004reproducible, tong2018revealing}. In Fig. \ref{fig:localcorrelation}c, we color each particle according to $D_i$ (again taken at $\delta t = 500\tau$), and indeed reveal a striking correlation between $\ntet$ and $D_i$. Intriguingly, both quantities show heterogeneity over approximately the same length scale. Clearly, local tetrahedrality is an excellent predictor for the dynamics of a  particle in the near future. 

This correlation can be quantified explicitly by calculating the Spearman's rank-order correlation between $\ntet(i)$ and $D_i$. In Fig. \ref{fig:localcorrelation}b, we plot this correlation as a function of the time interval $\delta t$ for both the small and large particles. At very short timescales ($\delta t \lesssim 0.1 \tau$), before any particles escape their cages, there is little correlation between local TLS and dynamic propensity. At intermediate timescales, we find a strong negative correlation between $\ntet(i)$ and $D_i$, confirming that  particles involved in fewer tetrahedra have higher mobility. The level of correlation attained by TLS outperforms most of the local observables  investigated so far \cite{tanaka2019revealing}. Finally, for timescales longer than the diffusion time $\tau_D \approx 10^4$, the memory of the initial configuration is lost, and the correlations start decaying back to zero.
An examination of these correlations in other hard-sphere mixtures provides similar results, with weaker correlations for systems with higher diffusivity (see SI). In faster systems, dynamics are less heterogeneous, and hence less predictable based on local structure.

The results in this Letter paint an elegant and comprehensive picture of the relationship between local structure and dynamics in hard-sphere mixtures, both on the global and local level. On the local level, examining tetrahedrality reveals large-scale regions of particles involved in a large number of tetrahedra, corresponding directly to areas of low mobility. Globally, the tetrahedrality of a hard-sphere mixture directly predicts its diffusivity, resulting in a data collapse of a vast variety of hard-sphere mixtures onto the same exponential curve. These predictions can be directly tested in experimental realizations of colloidal hard-sphere mixtures, and are strongly reminiscent of theoretical descriptions of glasses in terms of activation energies for collective rearrangement \cite{adam1965temperature}, and random first-order transition theory \cite{kirkpatrick1989scaling,bouchaud2004adam, berthier2011theoretical}. Most importantly, our results demonstrate that the dynamics of hard-sphere mixtures can be understood purely by looking for the most fundamental three-dimensional building block for the fluid: the simple tetrahedron.



\section{Methods}

\footnotesize

We perform standard EDMD simulations \cite{rapaport} of mixtures of hard spheres in the microcanonical ensemble (i.e. at constant number of particles $N$, volume $V$, and energy $E$). All particles have identical mass $m$. The time unit of the simulation is given by $\tau = \sqrt{\beta m \sigma^2}$, with $\sigma$ the diameter of the large spheres (for binary mixtures), or the average sphere size (for polydisperse mixtures), and $\beta = 1/k_B T$ with $k_B$ Boltzmann's constant and $T$ the temperature.

For the binary mixtures, we vary the size ratio $q = \sigma_S / \sigma_L$ (with $\sigma_{S(L)}$ the diameter of the small (large) spheres), as well as the fraction of large spheres $x_L$ and the packing fraction $\eta$. In particular, we explore size ratios $0.6 \leq q \leq 0.85$, compositions $0.2 \leq x_L \leq 0.65$, and packing fractions $0.5 \leq \eta \leq 0.58$. Systems which crystallized were excluded from all analysis. All systems contained at least $N=700$ particles, and measurements of local correlations in the slower systems were done using $N=2000$ particles. For the polydisperse systems, we used mixtures of spheres of 15 different equally spaced diameters, with the number of spheres of each diameter chosen from a Gaussian distribution with standard deviation $0.08 \leq s/\sigma \leq 0.20$.

In order to characterize the structure of the system, we find all tetrahedral clusters, defined as groups of four particles which are all considered nearest neighbors of each other. Nearest neighbors are determined via the same modified Voronoi construction used in the Topological Cluster Classification (TCC) algorithm \cite{malins2013tcc}, which differs from standard Voronoi constructions in two ways. First, two particles are only considered nearest neighbors if the line connecting them passes through the facet of the Voronoi cells shared by the two particles. Second, a cutoff parameter $f_c = 0.82$ is used to improve detection of four-membered rings of particles. Both modifications result in fewer connections between particles that are further apart than the typical first neighbor shell. For more details, see Ref. \cite{malins2013tcc}. Tetrahedra are then detected directly from the set of nearest-neighbor bonds, and each particle $i$ is labeled with the number $\ntet(i)$ of tetrahedra it is involved in. The global number of tetrahedra per particle is then defined as
\begin{equation}
   \langle \ntet \rangle = \frac{1}{N} \sum_i \ntet(i).
\end{equation}

In addition to locating tetrahedral clusters, we also measure the fraction of particles involved in icosahedral clusters in our system. For this, we use the TCC algorithm \cite{malins2013tcc} directly, again with a cutoff parameter $f_c = 0.82$.

Global dynamics are characterized via the diffusion time $\tau_D$. For binary mixtures, $\tau_D = \sigma_L^2 / D_\mathrm{bin}$, with $D_\mathrm{bin}$ the long-time self-diffusion coefficient averaged over all particles. For the polydisperse systems, $\tau_D = \sigma^2 / D_\mathrm{poly}$, with $D_\mathrm{poly}$ the long-time self-diffusion coefficient of particles with a diameter equal to the average diameter $\sigma$. Note that averaging over the diffusion of all species results in slightly lower diffusion times, due to the strong contribution from the highly mobile smallest species.

To characterize the dynamics on a per-particle basis, we measure the absolute displacement $\delta r_i (\delta t)$ for each particle $i$ after a time interval $\delta t$. As this quantity fluctuates heavily between different runs starting from the same configuration, we also make use of the dynamic propensity $D_i(\delta t)$, which is defined as the average value of $\delta r_i$ over many ($\sim 200$) runs starting from the same configuration but with randomized particle velocities. To quantify the correlation between $\ntet(i)$ and either $\delta r_i$ or $D_i$, we use Spearman's rank correlation coefficient, which measures how well the relation between two sets of data can be described by a monotonically increasing or decreasing function.

\bibliography{tetra}

\section*{Acknowledgements}
The authors thank Robert Botet, Paddy Royall, and Laura Filion for useful discussions. S.M.A. acknowledges CONACyT for funding.

\end{document}